\documentclass[lettersize,journal]{IEEEtran}
\usepackage[utf8]{inputenc} 
\usepackage[T1]{fontenc} 
\usepackage{breqn}
\usepackage{breqn}
\usepackage{flexisym}
\usepackage{mathtools}
\usepackage{multirow}
\usepackage{amsthm}
\usepackage{float}
\usepackage{placeins}
\usepackage{amsmath,amsfonts}
\usepackage{algorithmic}
\usepackage{algorithm}
\usepackage{array}
\usepackage[caption=false,font=normalsize,labelfont=sf,textfont=sf]{subfig}
\usepackage{epsfig}
\usepackage{epstopdf}
\usepackage{textcomp}
\usepackage{stfloats}
\usepackage{url}
\usepackage{verbatim}
\usepackage{graphicx}
\usepackage{cite}
\usepackage{xcolor}
\usepackage{multicol}
\newtheorem{theorem}{Theorem}

\begin{document}


\title{Achieving Hiding and Smart Anti-Jamming Communication: A Parallel DRL Approach against Moving Reactive Jammer}

\author{Yangyang Li, Yuhua Xu, Wen Li, Guoxin Li, Zhibing Feng, Songyi Liu, Jiatao Du, Xinran Li

\thanks{The authors are with the College of Communications Engineering, Army Engineering University of PLA, Nanjing 210000, China }
\thanks{This work is supported in part by the National Science Fund of China
for the Excellent Young Scholars under Grant 62122094, the National Natural Science Foundation of China under Grant 62071488, Grant U22B2002, and Grant 62201581~(Corresponding author: Yuhua Xu.)}

%

}


\maketitle

\begin{abstract}
This paper addresses the challenge of anti-jamming in moving reactive jamming scenarios. The moving reactive jammer initiates high-power tracking jamming upon detecting any transmission activity, and when unable to detect a signal, resorts to indiscriminate jamming. This presents dual imperatives: maintaining hiding  to avoid the jammer's detection and simultaneously evading indiscriminate jamming. Spread spectrum techniques effectively reduce transmitting power to elude detection but fall short in countering indiscriminate jamming. Conversely, changing communication frequencies can help evade indiscriminate jamming but makes the transmission vulnerable to tracking jamming without spread spectrum techniques to remain hidden. Current methodologies struggle with the complexity of simultaneously optimizing these two requirements due to the expansive joint action spaces and the dynamics of moving reactive jammers. To address these challenges, we propose a parallelized deep reinforcement learning (DRL) strategy. The approach includes a parallelized network architecture designed to decompose the action space. A parallel exploration-exploitation selection mechanism replaces the $\varepsilon $-greedy mechanism, accelerating convergence. Simulations demonstrate a nearly 90\% increase in normalized throughput.
\end{abstract}


\section{Introduction}
\IEEEPARstart{I}{n} wireless communication systems, the escalating threat posed by moving reactive jammers necessitates the development of advanced countermeasures~\cite{9733393,9449830}. Many existing works focus on the frequency domain, utilizing rapid frequency switching to evade various types of jamming. Evasion through the frequency domain is effective when the jammer's tracking capabilities are weak. However, as jamming techniques have advanced~\cite{10226352,10068300}, jammers now detect and track communication signals to release jamming, making frequency-domain anti-jamming less effective. These jammers release tracking jamming as soon as they detect any communication signals. Consequently, merely altering the frequency is not enough to handle such scenarios, leading to disrupted communications.

\textcolor[rgb]{0000,0.00,0000}{To effectively tackle these challenges, our approach commences with an in-depth look at optimizing frequency selection and spread spectrum parameters. For anti-jamming frequency selection, game-theoretic frameworks are commonly employed~\cite{6117758,4215896,8374072,7076591}. These methods model the interaction between the communication system and the jammer as strategic games, such as Stackelberg and Stochastic games. The Stackelberg game approaches provide structured frameworks for analyzing strategic interactions between the legitimate communication system and the jammer. The Stackelberg game model captures hierarchical decision-making processes, where one player (the leader) moves first, and the other (the follower) responds optimally. Conversely, the stochastic game framework accommodates scenarios with probabilistic transitions and repeated interactions, offering a dynamic perspective on the evolving strategies of both parties. Moreover, the game-theoretic approaches require knowledge about the jammer's behavior and power range~\cite{9777258,10734333}. While effective, these approaches are limited by their reliance on specific jammer information, highlighting the need for adaptive strategies for unknown moving reactive jammer.}

Spread spectrum technology can effectively lower the power density of a signal~\cite{Yang:24} thereby evading the jammer's detection. However, for certain reactive jammers~\cite{10416893}, when they do not detect a signal, they release indiscriminate jamming, meaning that simply relying on spread spectrum is not enough to effectively evade such jamming. Thus, joint optimization of spread spectrum and frequency agility is required to effectively counteract such jamming methods. Moreover, some reactive jammers are mobile, which further increases the difficulty of combating jammers.

Simultaneously optimizing spread spectrum and frequency agility is a good solution for countering jammers. However, this solution is fraught with its own set of difficulties. Namely, the curse of dimensionality~\cite{9904958}, a term used to describe the exponential increase in selectable actions can lead to suboptimal performance or even failure of the algorithm to converge towards a solution. This phenomenon poses a significant barrier to the practical application of multi-variable decision-making.
Secondly, for combating the mobility of jammers, deep Q-network (DQN) is commonly used. But for anti-jamming communication systems employing DQN-based algorithms~\cite{8314744,9105045,10287624,9613616,9264659,10227374}, the presence of the $\varepsilon $-greedy mechanism further complicates the situation~\cite{NEURIPS2022_2119b5ac}. The $\varepsilon $-greedy strategy is a balancing act between exploration~\cite{LADOSZ20221} and exploitation, which is crucial for learning optimal policies. In simple terms, it involves a trade-off: at times, the algorithm will choose to explore new actions (exploration), and at others, it will opt for actions based on what it currently deems best (exploitation).  However, the process of exploration-indispensable for the algorithm's learning inevitably slows down the rate at which the algorithm converges to an optimal policy. For communications, the requirement for instantaneous response is fundamental. Slow convergence impedes the fluidity of data flow, compromising the efficacy of message dissemination.

To tackle these challenges, we propose a parallelized \textcolor[rgb]{0000,0.00,0000}{deep reinforcement learning (DRL) scheme. }The proposed solution leverages a network architecture comprising a convolutional neural network~\cite{9451544} (CNN) and two fully connected networks dedicated to frequency and spreading factor selection. Through the utilization of dual neural networks, the vast action space is effectively decomposed, streamlining the decision-making process. Moreover, to address the slowness caused by the $\varepsilon$-greedy mechanism, the initial randomness of the training network is leveraged for exploration. As the applying network updates and refines its learning, enabling it to make more accurate selections. The instances where the applying network does not select the optimal action also contribute to the learning experience. By doing so, we circumvent the necessity of having a fixed number of iterations with randomness as mandated by the traditional $\varepsilon$-greedy mechanism.

 To conclude, our method significantly outperforms traditional DQN algorithms and hybrid approaches where frequency and spreading factor controls are separately optimized. In simulations highlight the superiority of our method, evidencing marked improvements in system about 90\% in normalized throughput. The computational requirements of the proposed algorithm have been analyzed, demonstrating its compatibility with a wide range of hardware configurations~\cite{10032057}. The main contributions of this work can be summarized as follows:
\begin{enumerate}
  \item \textbf{Novel parallelized DRL scheme for anti-jamming communications:} We introduce a parallelized DRL framework specifically tailored to combat reactive jamming in wireless communications. This scheme employs a parallel network architecture consisting of a CNN and two fully connected networks specialized for frequency and spreading factor selection, respectively. This innovative approach decomposes the complex action space into more manageable components, accelerating convergence and improving the efficacy of anti-jamming strategies.
  \item \textbf{Interconnected reward structure for enhanced coordination:} A strategic reward function is designed to ensure interdependency between the decisions made by the frequency and spreading factor selection networks. This interconnected reward structure promotes coherence and efficiency in the overall decision-making process, optimizing actions taken against sophisticated jamming tactics.
  \item \textbf{Design efficient exploration and exploitation mechanism:} The training dynamics of DRL algorithms are harnessed to implicitly balance exploration and exploitation, eliminating the need for explicit $\varepsilon$-greedy parameter tuning. This allows the algorithm to dynamically adjust its behavior based on the current training state, achieving faster convergence and improved adaptability in rapidly changing environments.
\end{enumerate}

\subsection{Related Works}
Many researchers have provided pertinent studies concerning the selection of spreading factor and frequency for anti-jamming. In~\cite{9380308,6117758,9050457} game theory is used to model jamming, but in our scenario, the dynamics of the jammer is unable to use game theory to model. In~\cite{7996380} proposes a control mechanism for adjusting the spreading factor in accordance with the distance between communicating nodes, aiming to optimize signal integrity under varying propagation distances, the control of spreading factor is used as the compared algorithm in simulation. \textcolor[rgb]{0000,0.00,0000}{Meanwhile,~\cite{7952524} introduces a two-dimensional anti-jamming decision framework, albeit limited to a binary choice of either exiting or remaining within a jammed zone, thus constraining the scope of adaptive responses, the algorithm is DQN, which was utilized and also considered as one of the baseline algorithm in our simulation for comparative analysis. In~\cite{10226268}, a reinforcement learning method is used for hopping pattern selection instead of frequency selection.}

In~\cite{8254362}, an anti-jamming underwater transmission framework that applies reinforcement learning to control the transmit power and uses the transducer mobility to address jamming in underwater acoustic networks. In \cite{8254362}, anti-jamming framework is proposed for underwater acoustic networks that utilizes reinforcement learning to optimize transmit power and leverages transducer mobility to mitigate jamming. In simulation, we used two channel model to test our algorithm.

The aforementioned papers either address issues within a single dimension using one method or, when dealing with multiple dimensions, still employ the DQN algorithm without decomposing the problem. Paper~\cite{9751039} decomposes anti-jamming decisions into parallel processes, yet overlooks the incorporation of spread spectrum techniques, a critical component for anti-jamming. Moreover, our proposed methodology eliminates the $\varepsilon$-greedy mechanism, thereby expediting the convergence of our learning algorithm. In comparison with~\cite{9837014}, which employs a deep deterministic policy gradient (DDPG) framework for real-time power control and unlike the scenario delineated in the referenced paper, where k-means~\cite{electronics9081295} clustering is employed to tackle positioning problems in an alternate dimension. Considering this idea of separate processing in~\cite{electronics9081295}, in our simulation, we have designed a comparative algorithm where the DQN handles frequency agility, while adaptive control~\cite{10120634} is in charge of the spreading factor adjustment.

\section{System Model and Problem Formulation}\label{sec:sys}
\begin{figure}[htpb]
\begin{center}
  \vspace{-1mm}
  \includegraphics[width=3.5in]{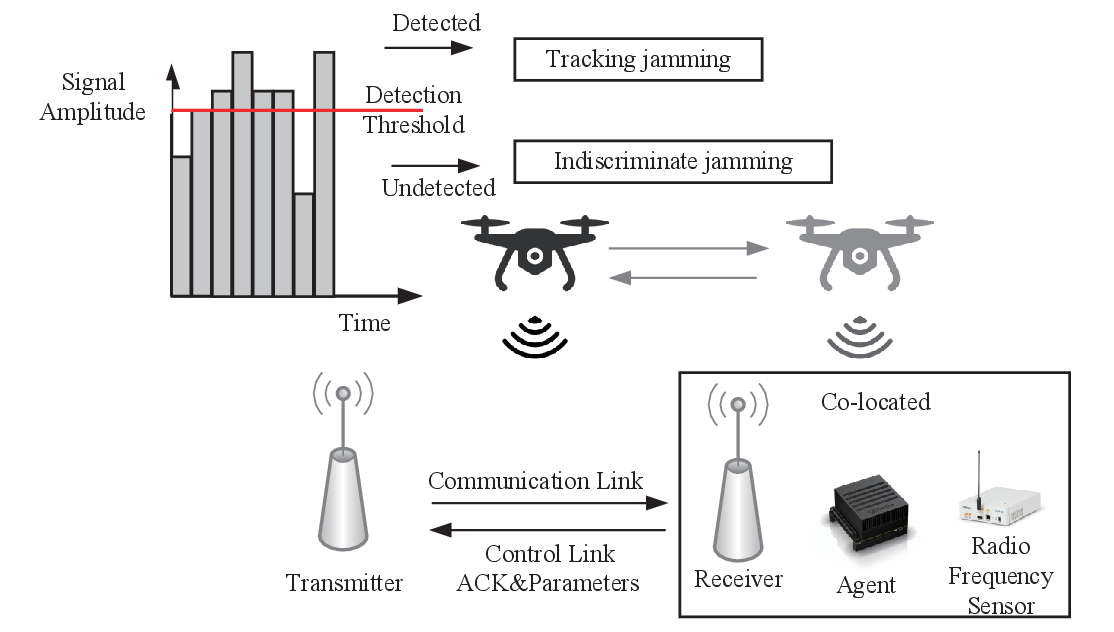}
  \caption{System model.}\label{Fig:sysmodel}
\end{center}
\end{figure}
\subsection{System Model}
As shown in Fig.~\ref{Fig:sysmodel}, we consider a classical communication scenario consisting of a pair of transceivers. The transmitter is able to adjust the spreading factor and communication frequency. The receiver, radio frequency sensor, and agent are co-located. The radio frequency sensor detects spectrum data, while the agent analyzes this data. Subsequently, the receiver, upon obtaining the analysis results from the agent, transmits the next communication parameters along with an acknowledgment (ACK) to the transmitter through a control link. The pair of transceivers is under reactive jamming by a mobile unmanned aerial vehicle (UAV) jammer. \textcolor[rgb]{0000,0.00,0000}{The moving reactive jammer} unleashes high-power tracking jamming once it detects activity from the transmitter. If the jammer fails to capture the user's signal, it proceeds to implement indiscriminate jamming, such as sweeping or comb jamming.

To be specific, for the transmitter, the available frequency range is defined as $[f_L, f_U]$, where $f_L$ and $f_U$ represent the lower and upper limit frequencies, respectively. This range is divided into $N$ channels, forming the set $F = \{f_1, f_2, ..., f_N\}$, with each $f_i$ denoting the center frequency of the $i$-th channel. The adjustable spreading factors for the transmitter are defined as $Q = \{ q_1, q_2, ..., q_M \}$, with each $q_i$ denoting a different spreading factor. Assuming the bandwidth used by the transmitter without spreading is $b_u$, the bandwidth utilized under the spreading factor $q_1$ will be $b_{\text{spread}}^1 = b_u \times q_1$. \textcolor[rgb]{0000,0.00,0000}{Assuming the PSD of transmission power} without spreading is $u(f)$, the PSD under the spreading factor $q_1$ will be $u_{\text{spread}}^1(f) = \frac{u(f)}{q_1}$.

For the radio spectrum sensor, at moment $t$, the sensor generates a spectrum waterfall ${S_t}$ containing the spectrum usage over the past $T$ time units and across a spectrum width $F$, which can be expressed as:

\begin{equation}\label{eq:1}
{S_t} = \left[ \begin{array}{*{20}{c}}
{s_{1}^{t - (N_T - 1)\Delta t}} & {s_{2}^{t - (N_T - 1)\Delta t}} & {...} & {s_{N_F}^{t - (N_T - 1)\Delta t}} \\
{s_{1}^{t - (N_T - 2)\Delta t}} & {...} & {...} & {...} \\
{...} & {...} & {...} & {...} \\
{s_{1}^t} & {s_{2}^t} & {...} & {s_{N_F}^t}
\end{array} \right],
\end{equation}
where $N_T = \frac{T}{\Delta t}$ and $N_F = \frac{F}{\Delta f}$ denote the sampling numbers in the time and frequency domains, respectively. $\Delta t$ and $\Delta f$ represent the time and frequency sampling resolutions, respectively. \textcolor[rgb]{0000,0.00,0000}{Therefore, the size of the spectrum waterfall $S_t$ is $N_T \times N_F$.} The discrete spectrum sample points in the sensed frequency spectrum matrix ${S_t}$ are expressed as:

\begin{equation}\label{eq:0}
s_{i}^{t - i\Delta t} = 10\log \left[\int_{i\Delta f}^{(i + 1)\Delta f} (n(f) + g_j j(f) + g_u u_{\text{spread}}(f)) \, df \right],
\end{equation}
where $n(f)$, $j(f)$, and $u(f)$ denote the \textcolor[rgb]{0000,0.00,0000}{PSD} of the sensed natural noise, the jammer's signal, and the transmitter's signal, respectively, $g_u$ and $g_j$ denote the channel gains of the transmitter and jammer, respectively.

For the jammer, we assume that the UAV jammer is able to launch high-power tracking jamming quickly once it detects the activities of the transmitter. When it fails to detect the user's signal, indiscriminate jamming (e.g., sweep or comb jamming) is launched to prevent the user from accessing the frequency spectrum. The detection threshold is $\alpha_{\text{th}}$. The received power at the mobile UAV jammer is $\alpha_t = g_{uj} p_t + \sigma^2$, where $p_t$ is the highest signal amplitude of the user, $g_{uj}$ denotes the channel gain from the user to the jammer, and $\sigma^2$ denotes the background noise. If the signal received by the jammer exceeds the threshold value $\alpha_t \ge \alpha_{\text{th}}$, it initiates tracking jamming; otherwise, it switches to indiscriminate jamming.

For the receiver, the signal-to-jamming-and-noise ratio (SJNR) is:

\begin{equation}
SJNR(f, q) = \frac{g_u g_{\text{sf}} \int_{f - b_{\text{spread}}/2}^{f + b_{\text{spread}}/2} u_{\text{spread}}(f) \, df}{\int_{f - b_{\text{spread}}/2}^{f + b_{\text{spread}}/2} (n(f) + g_j j(f)) \, df},
\end{equation}
where $g_{\text{sf}}$ is the processing gain in the spread frequency spectrum, and $g_u$ is the channel gain from the transmitter to the receiver. Assuming the demodulation threshold is $\beta_{\text{th}}$, the goal of our algorithm is to find the appropriate frequency and spreading factor to achieve $SJNR(f) \ge \beta_{\text{th}}$. The transmission rate is calculated as:

\begin{equation}\label{eq:trans}
C_t = \begin{cases}
b_u \cdot \log_2(1 + SJNR) & \text{if } SJNR \ge \beta_{\text{th}} \\
0 & \text{if } SJNR < \beta_{\text{th}}
\end{cases},
\end{equation}
where $C_t$ is calculated according to Shannon's equation~\cite{6773024} for information that meets the demodulation threshold.

\subsection{Problem Formulation}

In \textcolor[rgb]{0000,0.00,0000}{DRL}, the selection of anti-jamming decisions is conceptualized as a Markov process. Although the decision strategies may vary in our specific case, we shall later demonstrate that this process indeed adheres to the Markovian framework, and techniques such as DRL are aptly suited to address Markov Decision Process (MDP) problems. The objective of DRL-related algorithms is to maximize expected rewards through strategic optimization. Subsequently, we will formulate the problem and illustrate that this process conforms to the Markov decision framework.

The problem is generally formulated as tuples $\langle \mathbf{S, A, P, R} \rangle$, including state set $\mathbf{S}$, action set $\mathbf{A}$, transition probability set $\mathbf{P}$, and reward set $\mathbf{R}$.

\textbf{State:} The states in our case are pictures of spectrum waterfalls $\mathbf{S} = \{ S_1, S_2, \ldots, S_t, \ldots \}$, where these are obtained by the spectrum sensor.

\textbf{Action:} The actions of the transmitter are combinations of available frequency points $F = \{ f_1, f_2, \ldots, f_N \}$ and available spreading factors $Q = \{ q_1, q_2, \ldots, q_M \}$. The specific values of $M$ and $N$ are determined by the capabilities of the transmitter. Note, the actual decision-making process involves a combination of frequency and spreading parameters. Specifically, if frequency $f_1$ is chosen, then the compatible spreading factors available for selection are $Q = \{ q_1, q_2, \ldots, q_M \}$. Consequently, for each of the $N$ possible frequencies, there are $M$ associated spreading factor options. This pairing leads to an action space $\mathbf{A} = \{ (f_1, q_1), (f_1, q_2), \ldots, (f_1, q_M), (f_2, q_1), \ldots, (f_N, q_M) \}$, where the set $\mathbf{A}$ contains $M \times N$ elements.

\textbf{Transition probability:} The elements in $\mathbf{P}$ are state transition probabilities which can be expressed as
\begin{equation}
P(s_{t + 1}^p | s_t^p, f_t^p, q_t^p) = p(s_{t + 1}^p = S_{t + 1} | s_t^p = S_t, f_t^p = f_t, q_t^p = q_t),
\end{equation}
where $s_t^p$ denotes a possible frequency spectrum state, and $s_{t + 1}^p$ represents the subsequent state. The variables $f_t^p$ and $ q_t^p$ denote possible actions, whereas $S_t$ signifies the deterministic spectrum state, and $f_t$ and $ q_t$ are the corresponding deterministic actions. Specifically, the transition to the next deterministic spectrum state $S_{t + 1}$ is governed by the probabilistic distribution $S_{t + 1} \sim p(S_{t + 1} | S_t, f_t, q_t)$. The equation $P(s_{t + 1}^p | s_t^p, f_t^p, q_t^p) = p(s_{t + 1}^p = S_{t + 1} | s_t^p = S_t, f_t^p = f_t, q_t^p = q_t)$ defines the probability of executing an action \( (f_t, q_t) \) at time \( t \), leading to a transition into the next spectrum state \( S_{t + 1} \).


\textbf{Reward:} The reward set $\mathbf{R} = \{ r_1, \ldots, r_t, r_{t + 1}, \ldots \}$ contains the reward \textcolor[rgb]{0000,0.00,0000}{values, }where $r_t = R(S_t, f_t, q_t)$ represents the reward value $r_t$ obtained after making an action $(f_t, q_t)$ in the state $S_t$. Given the initial spectrum state distribution $\rho_1(S_1)$, the occurrence probability of a $T$-step action trajectory in the Markov decision process is:

\begin{equation}\label{eq:3}
p(\tau | \pi) = \rho_1(S_1) \prod_{t = 1}^{T - 1} p(S_{t + 1} | S_t, f_t, q_t) \pi(f_t, q_t | S_t),
\end{equation}
where $\tau$ is the action trajectory and $\pi(f_t, q_t | S_t)$ denotes the probability of selecting $(f_t, q_t)$ in state $S_t$. Therefore, the expected reward function is:

\begin{equation}\label{eq:4}
J(\pi) = \mathbb{E}_{\tau \sim \pi}[R(\tau)].
\end{equation}

Then, the aim of the optimal policy $\pi^*$ is to find the best communication  frequency $f$ in every spectrum state, and it can be expressed as:
\begin{equation}\label{eq:5}
\pi^* = \arg \max_\pi J(\pi).
\end{equation}

\subsection{The proof of the Markov Decision Process}

\begin{theorem}
In our case, if the set of actions $\mathbf{A}$ employed aligns with the set of optimal actions $\mathbf{A}^*$, and if a sufficiently large retrospective value $H$ can be recorded, then this process is considered a Markov Decision Process.
\end{theorem}

\begin{IEEEproof}
To prove the process is a Markov process, one must show that $\Pr(S_{k + 1} \mid S_k, a_k)$, where $\Pr(\cdot)$ represents the probability, a measure of the likelihood of an event occurring on a scale from 0 to 1, means the probability depends only on the current state and the next state. Assume the historical environmental states are denoted as $S_1, S_2, S_3, \ldots, S_k$ and historical actions are denoted as $(f_1, q_1), (f_2, q_2), (f_3, q_3), \ldots, (f_k, q_k)$. According to the definition of state transition, the probability of the environmental state transitioning from $S_k$ to $S_{k + 1}$ is given by
\begin{equation}\label{eq:3}
\begin{array}{l}
\Pr(S_{k + 1} \mid S_k, f_k, q_k, S_{k - 1}, f_{k - 1}, q_{k - 1}, \ldots, S_0, f_0, q_0) \\
\stackrel{(a)}{=} \Pr(S_{k + 1} \mid S_k, a_k, S_{k - 1}, a_{k - 1}, \ldots, S_0, a_0) \\
\stackrel{(b)}{=} \Pr(o_k, o_{k - 1}, \ldots, o_{k - H + 1} \mid o_k, \ldots, o_0, a_k, \ldots, a_0) \\
= \Pr(o_{k + 1} \mid o_k, o_{k - 1}, \ldots, o_0, a_k, \ldots, a_0)
\end{array}
\end{equation}
where (a) comes from $a_k = (f_k, q_k)$, and in (b), $o_k$ is the observation of the frequency spectrum at time $k$, $H$ refers to the depth of observation. $o_{k+1}$ is the observation of the frequency spectrum at time $k+1$. In this context, the user's observed plan at iteration $k$ is only related to the decisions at the previous iterations and the large model itself. Furthermore, if $H$ is long enough, the current observed state is only dependent on the states within the relevant trajectory horizon $H_r$, and is independent of observation states beyond the trajectory horizon $H$. Due to the trajectory length of observations being greater than the relevant time length, the signal state is unrelated to the states beyond the retrospective length $H$. Therefore, equation~\ref{eq:3} is equivalent to:
\begin{equation}
\begin{array}{l}
\Pr(S_{k + 1} \mid S_k, f_k, q_k, S_{k - 1}, f_{k - 1}, q_{k - 1}, \ldots, S_1, f_1, q_1) \\
= \Pr(S_{k + 1} \mid S_k, a_k, S_{k - 1}, a_{k - 1}, \ldots, S_1, a_1) \\
= \Pr(o_{k + 1} \mid o_k, o_{k - 1}, \ldots, o_{k - H_r + 1}, a_k) \\
= \Pr(o_{k + 1} \mid o_k, o_{k - 1}, \ldots, o_{k - H + 1}, a_k) \\
= \Pr(o_{k + 1}, o_k, \ldots, o_{k - H + 2} \mid o_k, o_{k - 1}, \ldots, o_{k - H + 1}, a_k) \\
= \Pr(S_{k + 1} \mid S_k, a_k).
\end{array}
\end{equation}
\end{IEEEproof}

Therefore, when the retrospective length $H$ is sufficiently large, this problem satisfies the Markov property, making it a Markov Decision Problem (MDP). When the problem is an MDP, there exists a Markov equilibrium in the problem.



\section{A Parallel DRL Exploration Free Anti-jamming Strategy}\label{sec:alg}

\begin{figure*}[htpb]
\begin{center}
  \vspace{-1mm}
  \includegraphics[width=7in]{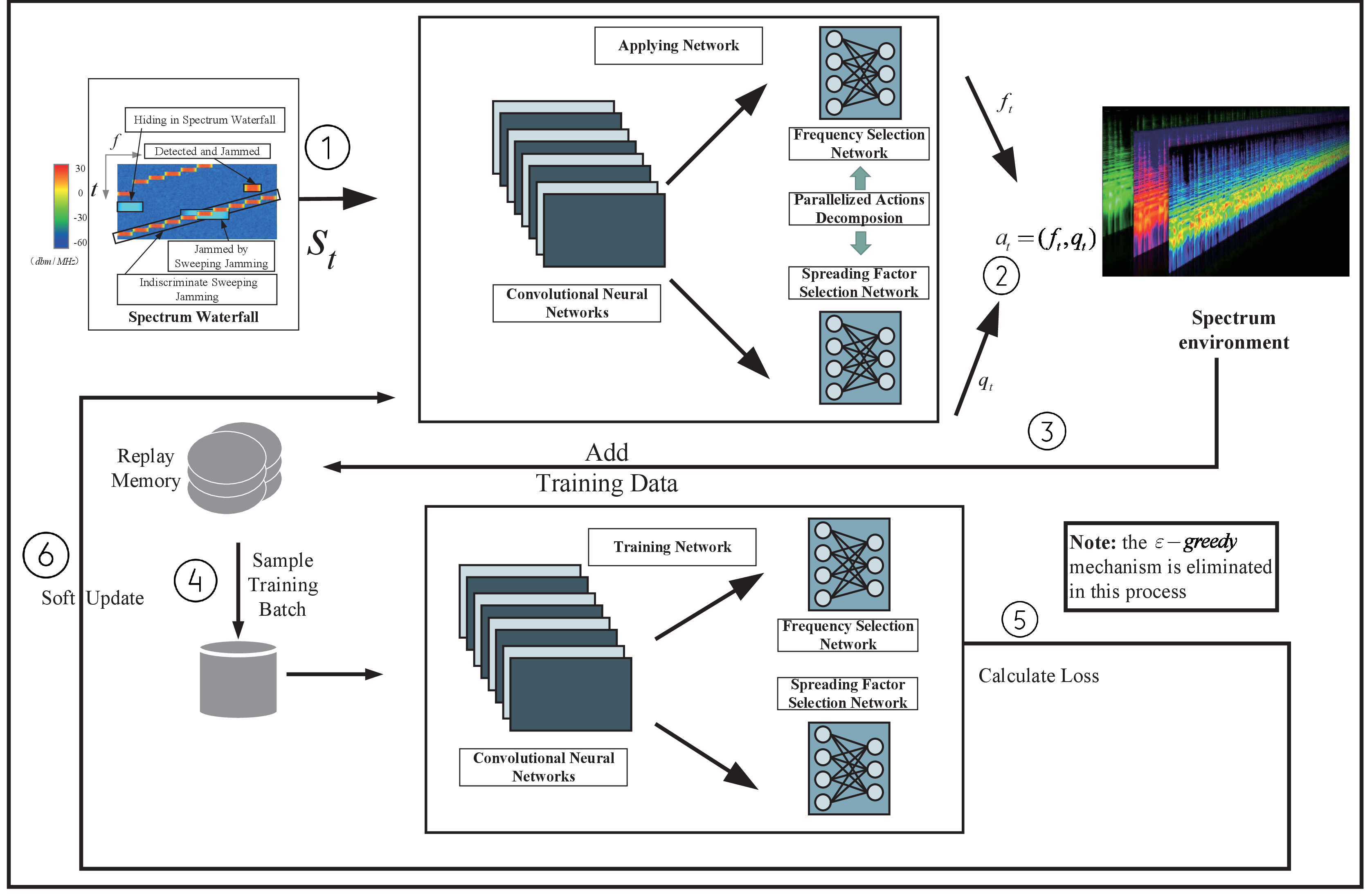}
  \caption{The illustration of the algorithm steps. }\label{Fig:Network}
\end{center}
\end{figure*}

 In this paper, the vast action space is partitioned into two parallel action spaces. Accordingly, corresponding reward functions are designated for individual actions, alongside a collaborative reward function. Moreover, tailored designs are implemented for the network architecture, and updating methods are designed for parallel actions. As shown in Fig~\ref{Fig:Network}, we introduce our algorithm in steps as follows: encompassing its inputs (step 1), outputs (step 2), network architecture, training process (steps 3, 4, 5), and updating approach (step 6).

\subsection{Network Architecture, Input and Output}

\textbf{Network Architecture:} Targeting frequency spectrum data and parallel actions, a novel architecture has been designed as shown in Fig~\ref{Fig:Network}, comprising a CNN coupled with two parallel fully connected networks (FCNs), diverging from the conventional DQN setup. Unlike the standard DQN where the CNN is sequentially followed by fully connected layers, this tailored configuration incorporates a split after the CNN stage. These twin FCNs operate in parallel: one is dedicated to the selection of communication frequencies, which is the frequency selection network, while the other is responsible for choosing communication spreading factors, which is the spreading factor selection network.

\textbf{Input:} The input to the network is a spectrum waterfall, which has been previously defined in the system model as $S_t$. As shown by the spectrum waterfall \textcolor[rgb]{0000,0.00,0000}{in Fig.~\ref{Fig:Network}, }step 1, if the transmission uses too small a spreading factor, it can be easily detected and tracked due to excessive power. If too large a spreading factor is chosen, the communication occupies a wider band, making it more susceptible to sweeping jamming. Therefore, only by selecting an appropriate spreading factor and an appropriate frequency point can successful communication be achieved.

\textbf{Output:} The output is step 2 in Fig~\ref{Fig:Network}. By decomposing the selection process of communication frequencies and spreading factors, the originally combined action space of ${{M}} \times N$ can be effectively fragmented into two separate and more manageable subspaces, one with $M$ for frequency actions and another with $N$ for spreading factor actions. This strategic separation significantly reduces the complexity of the action space, transforming it from a large, joint space requiring the evaluation of ${{M}} \times N$ possible combinations at every decision point, to two distinct spaces involving $M+N$ total actions.

\subsection{Training Process}

At the initialization stage of the training process, hyperparameters such as the learning rate, discount factor, and the size of the experience replay buffer need to be set. Following this, the other components are introduced.

\textbf{Replay Memory:} Replay memory involves storing the agent's past experiences (transitions) in a buffer and randomly sampling mini-batches from this buffer for training, which corresponds to steps 3 and 4 in Fig.~\ref{Fig:Network}. This technique breaks the correlation between consecutive samples and improves the stability of the training. In our case, the experience replay buffer stores ${e_{t+1}} = ({S_t},{f_t},{q_t},r_t^f,r_t^q,{s_{t + 1}})$, where $S_t$ and $S_{t+1}$ represent the spectrum waterfalls, $f_t$ and $q_t$ are the actions taken in the spectrum waterfall $S_t$. In the training of different networks, the data is used differently; for the frequency selection network, the training data is $({S_t},{f_t},r_t^f,{s_{t + 1}})$, and for the spreading factor selection network, the training data is $({S_t},{q_t},r_t^q,{s_{t + 1}})$.

\textbf{Reward Function Design:}
Next, we design the reward functions for the frequency selection network and the spreading factor selection network, respectively. The reward function for the frequency selection network is defined as:
\begin{equation}
r_t^f = \eta \cdot \delta(f,q),
\end{equation}
where $\eta$ represents a positive reward scaling factor, tasked with regulating the magnitude of rewards to remain within a reasonable interval, and $\delta(f,q)$ is an indicator function signifying the success of communication, defined as follows:
\begin{equation}
\delta(f,q) = \left\{ \begin{array}{l}
1,~~~SJNR \ge {\beta _{th}}\\
0,~~~SJNR < {\beta _{th}},
\end{array} \right.
\end{equation}
where $SJNR \ge {\beta _{th}}$ indicates the success of communication, resulting in $\delta(f,q)=1$, and $SJNR < {\beta _{th}}$ indicates the failure of communication, resulting in $\delta(f,q)=0$.
The reward function for the spreading factor selection network is defined as:
\begin{equation}
r_t^q = \eta \cdot C_t \cdot \delta(f,q) - \kappa \cdot \frac{1}{{b_{\text{spread}}}},
\end{equation}
where $C_t$ is the transmission rate defined in section~\ref{sec:sys}, $\kappa$ is the scale factor for $\frac{1}{{b_{\text{spread}}}}$, and $\kappa \cdot \frac{1}{{b_{\text{spread}}}}$ limits the spread spectrum bandwidth, reducing its occupancy of the frequency spectrum.

\textbf{Training Network and Applying Network:} To further enhance stability and apply our algorithm online, we use the training network and applying network in parallel. The training network takes actions in the environment and collects data for the replay memory, which is then used by the training network for training. After a certain number of iterations, the parameters of the applying network are updated based on the training network. Both networks approximate the Q-value function; for the training network, the function is
\begin{equation}
Q(S,a;\theta),
\end{equation}
where $\theta$ represents the trainable parameters of the neural network. The applying network maintains a stable estimate of the Q-values $Q(S,a;\theta^-)$, with parameters $\theta^-$.

\begin{figure}
\begin{center}
  \vspace{-1mm}
  \includegraphics[width=3.5in]{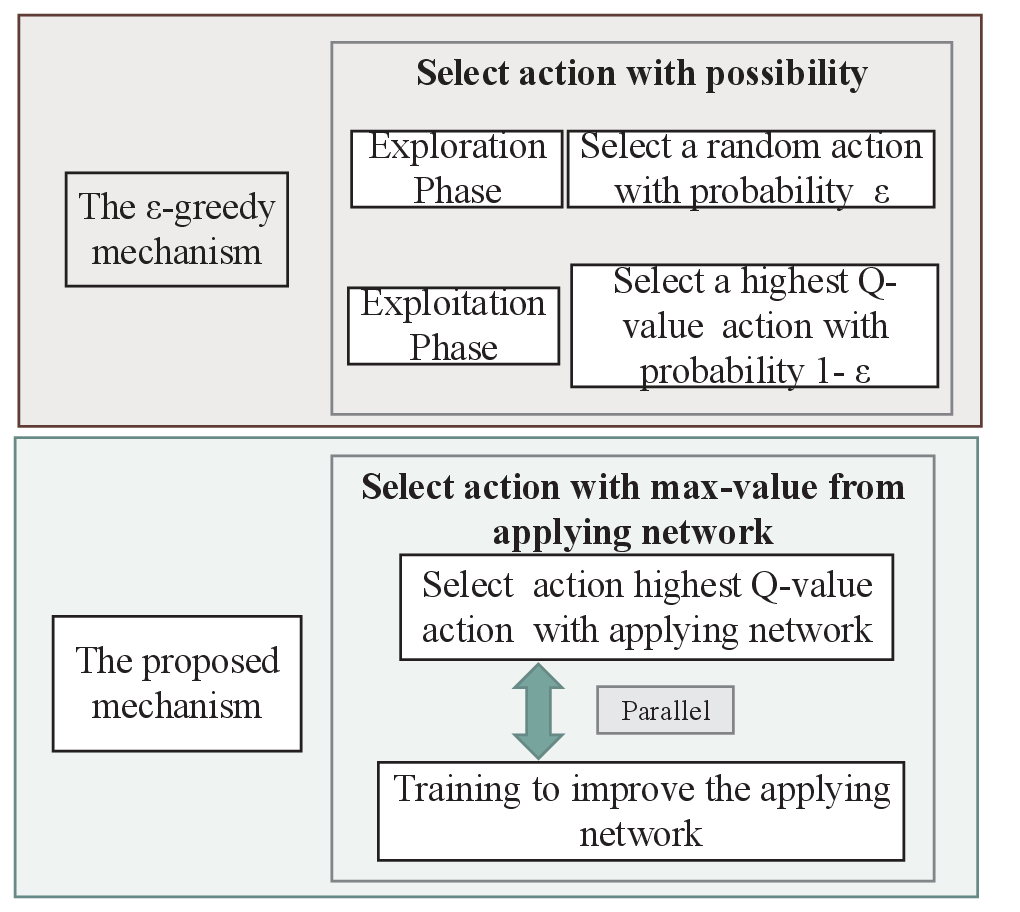}
  \caption{Comparison of the designed mechanism and the $\varepsilon$-greedy mechanism.}\label{Fig:Compare}
\end{center}
\end{figure}

To reduce the number of training iterations and make our algorithm online, the conventional $\varepsilon$-greedy mechanism has been omitted in our design. The $\varepsilon$-greedy mechanism is a fundamental component in DQN-based algorithms that strikes a balance between exploitation and exploration. It operates by choosing the action with the highest known value with probability $1 - \varepsilon$, thus exploiting current knowledge, and selecting any action randomly with probability $\varepsilon$, thereby exploring uncharted territory. Instead, we have substituted the randomness of the $\varepsilon$-greedy approach with the inherent randomness of the applying network.

Fig. \ref{Fig:Compare} shows the differences. The exploration-exploitation trade-off in the $\varepsilon$-greedy algorithm arises from an exploration rate ($\varepsilon$) that balances between exploring unknown actions and exploiting the current best action. A higher exploration rate implies that the algorithm is more inclined to attempt new, unevaluated actions, which can result in a failure to converge rapidly to the optimal solution in the short term. In our designed mechanism, the action is always selected as the maximum value from the applying network. Since the applying network directly selects the maximum value, it can reflect the current state of the network training in a real-time manner, without the need to wait for the probability of random selection to decrease. This aspect is helpful in improving the convergence speed.

However, this mechanism introduces instability into the training process. To address this potential issue, we have implemented a soft update design which is introduced in detail in the next subsection.

\textbf{Loss Function and Update:} The loss function quantifies the discrepancy between the predicted Q-values and the target Q-values. The calculation of the loss function is step 6 in Fig.~\ref{Fig:Network}. The update of the Q-values function is defined as:
\begin{equation}
Q\left( {{S_t},{a_t}} \right) \leftarrow (1 - \alpha)Q\left( {{S_t},{a_t}} \right) + \alpha ({r_{t + 1}} + \gamma \max_{{a^\prime}}Q\left( {{S_{t + 1}},{a^\prime}} \right)),
\end{equation}
where $Q\left( {{S_t},{a_t}} \right)$ represents the current estimate of the expected future reward for taking action ${a_t}$; the symbol $\leftarrow$ indicates assignment, meaning the value on the left is updated to the value calculated on the right; $\alpha$ is the learning rate, a scalar between 0 and 1, which determines the extent to which new information overrides old information. Smaller values imply slower learning, while larger values allow for quicker adaptation but risk instability; $r_{t+1}$ is the immediate reward obtained after performing an action in a state, providing direct feedback on the quality of the action; $\gamma$ is the discount factor, also a value between 0 and 1, which discounts future rewards relative to immediate ones. It balances the trade-off between immediate and long-term gains, with values closer to 1 giving more weight to future outcomes; $\gamma \max_{{a^\prime}}Q\left( {{S_{t + 1}},{a^\prime}} \right)$ is the maximum expected future reward achievable from the next state $S_{t+1}$ by taking any possible action. This term encourages the algorithm to select actions that lead to potentially higher rewards in the next states.

For the frequency selection network, the target value $y_t^f$ is defined as:
\begin{equation}
y_t^f = r_t^f + \gamma \max_{f'}Q\left( {{S_{t + 1}},f';\theta ^ - } \right),
\end{equation}
where $y_t^f$ combines the immediate reward $r_t^f$ with the estimated optimal future return. It serves as the goal for what the Q-value should be in the current state-action pair, encouraging the network to adjust its actions accordingly. Note that the estimation of the maximum expected future reward achievable from a state under any action $Q\left( {{S_{t + 1}},f';{\theta ^ - }} \right)$ is computed using the applying network parameters ${\theta ^ - }$. The use of a separate set of weights in the applying network ensures that the target values change more gradually than the estimates from the training network, contributing to learning stability.
Then, the loss function is defined as:
\begin{equation}
L_f(\theta)=\mathbb{E}_{\left(S, f, r^f, S^{\prime}\right)\sim U(D)}\left[\left(y_t^f-Q\left(S_t, f_t ; \theta\right)\right)^2\right],
\end{equation}
where the loss is defined as the expected value over tuples sampled uniformly $U(D)$ from the experience replay buffer $D$.

Similarly, for the spreading factor selection network, the target value $y_t^q$ is defined as:
\begin{equation}
y_t^q = r_t^q + \gamma \max_{q'}Q\left( {{S_{t + 1}},q';\theta ^ - } \right),
\end{equation}
and the loss function is defined as:
\begin{equation}
L_q(\theta)=\mathbb{E}_{\left(S, q, r^q, S^{\prime}\right)\sim U(D)}\left[\left(y_t^q-Q\left(S_t, q_t ; \theta\right)\right)^2\right].
\end{equation}
The network parameters are then updated through backpropagation by minimizing this loss using an optimizer, commonly Adam or RMSProp~\cite{choi2019empirical}, which adjusts the network weights to minimize the action error over $N_r$ iterations.
\begin{figure}[h]
\begin{center}
  \vspace{-1mm}
  \includegraphics[width=3.5in]{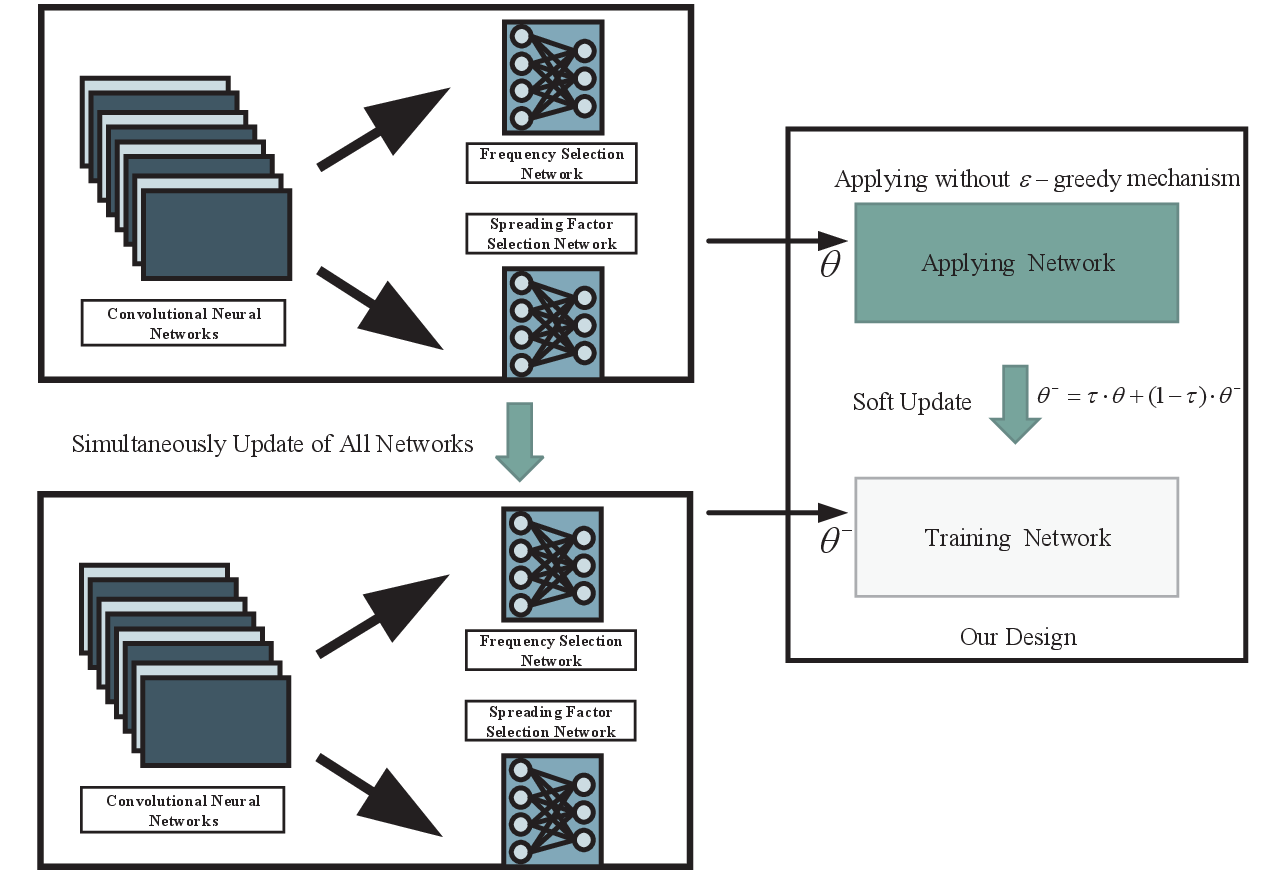}
  \caption{The simultaneously update of our design.}\label{Fig:update}
\end{center}
\end{figure}
\textcolor[rgb]{0000,0.00,0000}{Fig.~\ref{Fig:update}, shows the process of the update of the network, specifically, each update involves updating the entire set of network parameters, and we employ a soft update mechanism. Soft updates involve gradually changing the target network parameters by blending a small portion of the new parameters with the existing ones. This approach ensures stability and smooth transitions during the learning process~\cite{KOBAYASHI202163}, preventing abrupt changes that could destabilize the model.}
\begin{equation}\label{eq:update}
{\theta ^ - } = \tau  \cdot \theta  + (1 - \tau ) \cdot {\theta ^ - }
\end{equation}
\textcolor[rgb]{0000,0.00,0000}{where \( \tau \) is a small positive value, ensuring that updates are incremental and stable.}

\begin{algorithm}
	\caption{Parallelized DRL Exploration Free Anti-Jamming Algorithm}
	\label{alg:1}
    \emph{Initialization:} Training network parameters $\theta$, applying network parameters ${\theta ^ - }$, experience replay memory $D$, learning rate $\alpha $, update iteration $N_r$, soft update parameter $\tau$, scale factor $\kappa$\\
    1: Obtain initial state $S$ from the environment.  \\
   \textbf{Repeat}: $t = 0,\Delta t,2\Delta t,...,\infty $
	\begin{algorithmic}
        \STATE 2: Select the action with the highest Q-value according to the target Q-network.
        \STATE 3: Execute action $(f,q)$ and observe the reward $r^f$,$r^q$ and the new state $S'$.
        \STATE 4: Store the tuple $({S},{f},{q},r^f,r^q,{S'})$ in the experience replay memory $D$.
        \STATE 5: Sample a random mini-batch of experiences from $D$.
        \STATE 6: Calculate the loss function of the frequency selection network using $({S},{f},r^f,{S'})$
        \STATE 7: Calculate the loss function of the spreading factor selection network using $({S},{q},r^q,{S'})$
        \STATE 8: Periodically $N_r$ update the applying network parameters using soft update
	\end{algorithmic}
\end{algorithm}

\section{Simulation Results}
\subsection{Different scenarios for testing algorithm}

\begin{table}[]
\centering
\caption{}
\label{tab:spara}
\begin{tabular}{ll}
\hline
\textbf{Parameter}            & \textbf{Value}         \\ \hline
\textbf{UAV Jammer}           &                        \\
Patrol Initial Position       & $(0,0,300)$m           \\
Moving Speed                  & $v = 20m/s$            \\
Patrol End Position         & $(1000,1000,500)$m     \\
Jamming Power                 & $60dBm$                \\
Detection Threshold           & $-70dBm$               \\ \hline
\textbf{Transmitter Location} & $(0,0,0)$m             \\
User's Max Power              & $200mW$                \\
Spreading Factors             & 1,2,4,8,16,32          \\
Base Signal Band              & 0.5MHz                 \\ \hline
\textbf{Receiver Location}    & $(1000,0,0)$m          \\ \hline
Frequency Band                & 800-820MHz             \\
Number of Channels            & 10 (non-overlapping)   \\
Spectrum Sensing Interval     & $10ms$                 \\
Sampling Resolution           & $\Delta f = 0.1MHz$    \\ \hline
\textbf{Signal Waveform}      & Raised Cosine Waveform \\ \hline
Roll-off Factor               & $0.6$                  \\ \hline
\textbf{Channel Model} & \begin{tabular}[c]{@{}l@{}}1. Free Space Path Loss (FSPL) \\ 2. FSPL +Shadow Loss (4dB) \end{tabular} \\ \hline
\end{tabular}
\end{table}

\begin{figure}[htpb]
\begin{center}
  \vspace{-1mm}
  \includegraphics[width=3.5in]{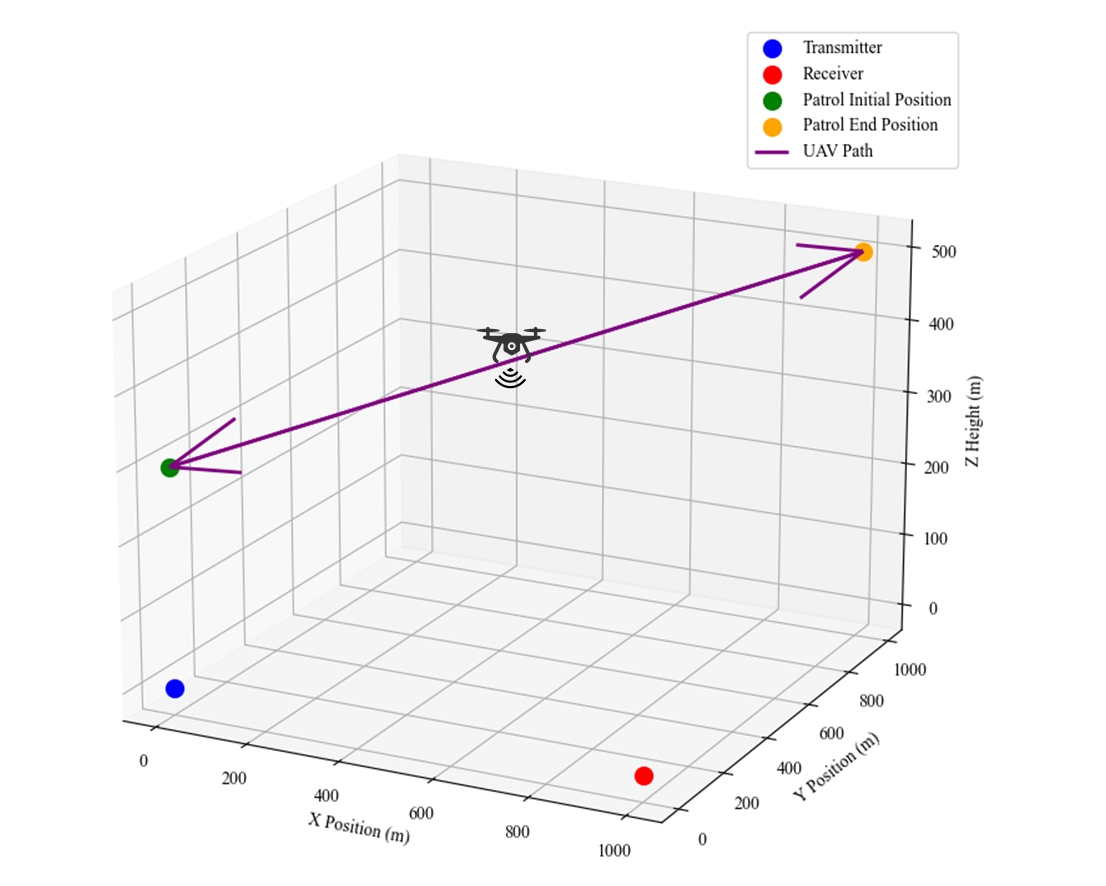}
  \caption{The locations of transceivers and patrol route of jammer }\label{Fig:Position}
\end{center}
\end{figure}

The typical scenario is considered, which includes a mobile UAV jammer located at (0,0,300)m initially, \textcolor[rgb]{0000,0.00,0000}{flying horizontally with a moving speed $v = 20$m/s and returning to (1000, 1000, 500)m. A transmitter is situated at (0, 0, 0)m and a receiver at (1000, 0, 0)m. In Fig.~\ref{Fig:Position}, the positions listed in Tab.~\ref{tab:spara} are marked, and the patrol route is shown in purple. The whole band 800-820MHz is divided into ten non-overlapping channels. Spectrum sensing is performed every $10$ms with a sampling resolution $\Delta f = 0.1$MHz.} A raised cosine waveform is used with a roll-off factor of 0.6. The maximum power of the user is 200mW, the power of the UAV jammer is 60dBm, and the detection threshold is -70dBm.

Two channel models are used in testing the designed algorithm. One is the free space path loss (FSPL) model~\cite{9586048}, which represents the loss of signal strength that naturally occurs when a radio wave propagates through free space, devoid of obstructions like buildings or terrain features. The FSPL model is defined as:
\begin{equation}
L_{FSPL} = 20\log_{10}\left(\frac{4\pi df}{c}\right) + 20\log_{10}(d),
\end{equation}
where $L_{FSPL}$ is the path loss in decibels (dB), $d$ is the distance between the transmitter and receiver (in meters), $f$ is the frequency of the signal (in Hertz), and $c$ is the speed of light in a vacuum (approximately \(3 \times 10^8\) m/s).

Another channel model combines the FSPL with shadowing fading~\cite{REN2011626,6882778} and\textcolor[rgb]{0000,0.00,0000}{ is commonly used to describe the propagation environment where large obstacles or buildings can significantly impact signal strength.} This combination allows for a more realistic depiction of signal attenuation over distance and through various environmental conditions to test our design. The combined FSPL and shadowing fading model can be mathematically represented by the following equation:
\begin{equation}
L_{F+S}=L_{FSPL}+X,
\end{equation}
\textcolor[rgb]{0000,0.00,0000}{where $X$ is the shadowing component in dB~\cite{6882778}, which follows a log-normal distribution characterized by its}~\textcolor[rgb]{0000,0.00,0000}{PSD}:
\begin{equation}
f(X;\mu ,\sigma ) = \frac{1}{{X\sigma \sqrt {2\pi } }}\exp \left( { - \frac{{{{\left( {\ln X - \mu } \right)}^2}}}{{2{\sigma ^2}}}} \right).
\end{equation}
where $\mu$ is the mean of the distribution in the natural logarithmic scale, and $\sigma$ is the standard deviation of the distribution in dB. In urban and rural environments, standard deviations range from 2.7 to 5.6 dB~\cite{4067118}, and a value of 4 dB is used in the simulation.

In Table~\ref{tab:spara}, other details are shown. For frequency hopping spread spectrum, the spread factor \textcolor[rgb]{0000,0.00,0000}{is the power of two,} and in many communication systems like widely used chirp spread spectrum (CSS)-based long-range (LoRa)~\cite{10403992}, the selection set of spreading factors is \{1, 2, 4, 8, 16, 32\}. \textcolor[rgb]{0000,0.00,0000}{To sum up, we primarily utilized a self-generated dataset through simulation to evaluate and validate the performance of our proposed anti-jamming algorithms, and parameters are from empirically measured.}

%

\subsection{Details for Networks}
\begin{table*}[]
\centering
\caption{The example hyperparameters for designed algorithm}
\label{tab:Parameters}
\begin{tabular}{|c|c|c|c|c|l|ll|}
\hline
\textbf{Network} &
  \textbf{Layer} &
  \textbf{Setting} &
  \textbf{Input Size} &
  \textbf{Output Size} &
  \multicolumn{1}{c|}{\textbf{\begin{tabular}[c]{@{}c@{}}Other\\ Hyperparameters\end{tabular}}} &
  \multicolumn{1}{c|}{\textbf{Description}} &
  \textbf{Value} \\ \hline
\multirow{2}{*}{\begin{tabular}[c]{@{}c@{}}Convolutional \\ Neural Networks\end{tabular}} &
  \begin{tabular}[c]{@{}c@{}}Convolutional\\ layer\end{tabular} &
  \begin{tabular}[c]{@{}c@{}}kernel_size=4, \\ stride=2\end{tabular} &
  200*200 &
  99*99 &
  $D$ &
  \multicolumn{1}{l|}{Size of replay memory} &
  1000 \\ \cline{2-8}
 &
  \begin{tabular}[c]{@{}c@{}}Convolutional\\ layer\end{tabular} &
  \begin{tabular}[c]{@{}c@{}}kernel_size=4, \\ stride=2\end{tabular} &
  99*99 &
  48*48 &
  $\alpha$ &
  \multicolumn{1}{l|}{Learning rate} &
  0.001 \\ \hline
\multirow{3}{*}{\begin{tabular}[c]{@{}c@{}}Frequency \\ Selection Network\end{tabular}} &
  \begin{tabular}[c]{@{}c@{}}Fully connect\\ layer\end{tabular} &
  Activation: ReLU &
  48*48 &
  512 &
  $N_r$ &
  \multicolumn{1}{l|}{Update iteration} &
  16 \\ \cline{2-8}
 &
  \begin{tabular}[c]{@{}c@{}}Fully connect\\ layer\end{tabular} &
  Activation: ReLU &
  512 &
  256 &
  $\tau$ &
  \multicolumn{1}{l|}{Soft update parameter} &
  0.2 \\ \cline{2-8}
 &
  \begin{tabular}[c]{@{}c@{}}Fully connect\\ layer\end{tabular} &
  Activation: ReLU &
  256 &
  \begin{tabular}[c]{@{}c@{}}Number of \\ available frequencies\end{tabular} &
  $\kappa$ &
  \multicolumn{1}{l|}{Reward Scale factor} &
  0.2 \\ \hline
\multirow{3}{*}{\begin{tabular}[c]{@{}c@{}}Spreading Factor \\ Selection Network\end{tabular}} &
  \begin{tabular}[c]{@{}c@{}}Fully connect\\ layer\end{tabular} &
  Activation: ReLU &
  48*48 &
  512 &
  $\eta$ &
  \multicolumn{1}{l|}{Reward Scale factor} &
  0.2 \\ \cline{2-8}
 &
  \begin{tabular}[c]{@{}c@{}}Fully connect\\ layer\end{tabular} &
  Activation: ReLU &
  512 &
  256 &
  Optimizer &
  \multicolumn{2}{c|}{Adam} \\ \cline{2-8}
 &
  \begin{tabular}[c]{@{}c@{}}Fully connect\\ layer\end{tabular} &
  Activation: ReLU &
  256 &
  \begin{tabular}[c]{@{}c@{}}Number of \\ spreading factors\end{tabular} &
  Loss &
  \multicolumn{2}{c|}{Mean Square Error} \\ \hline
\end{tabular}
\end{table*}

In this section, the network architecture and other hyperparameters are introduced. Hyperparameters are pre-set variables that govern the learning process and model architecture, fine-tuned to optimize performance, distinct from the model's learnable parameters. In summary, as mentioned in Section~\ref{sec:alg}, the entire network consists of three sub-networks: a convolutional neural network and two fully connected neural networks for decision-making, namely the frequency selection network and the spreading factor selection network. Their specific parameters are outlined in Table~\ref{tab:Parameters}. Additionally, other hyperparameters are also included in the table. Notably, hyperparameters are set before training and indirectly govern the learning process. For instance, these include the learning rate, scaling factor for reward values, and the configuration of the optimizer.

\subsection{Performance Comparison}
\begin{figure}[htpb]
\begin{center}
  \vspace{-1mm}
  \includegraphics[width=3.5in]{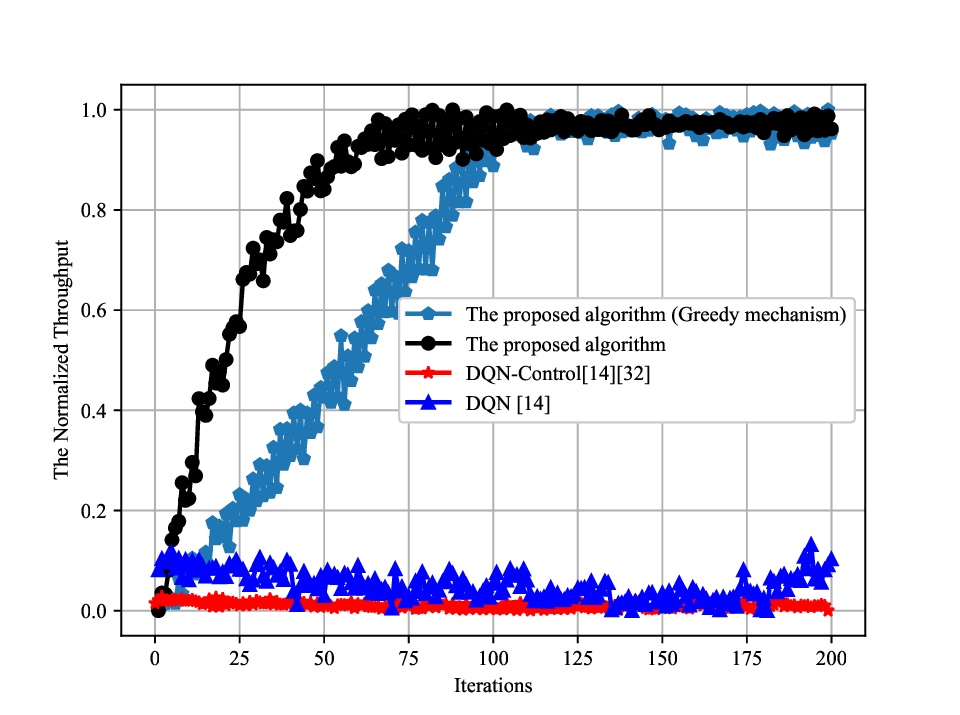}
  \caption{The performance of different algorithms.}\label{Fig:Result}
\end{center}
\end{figure}
\begin{figure}[htpb]
\begin{center}
  \vspace{-1mm}
  \includegraphics[width=3.5in]{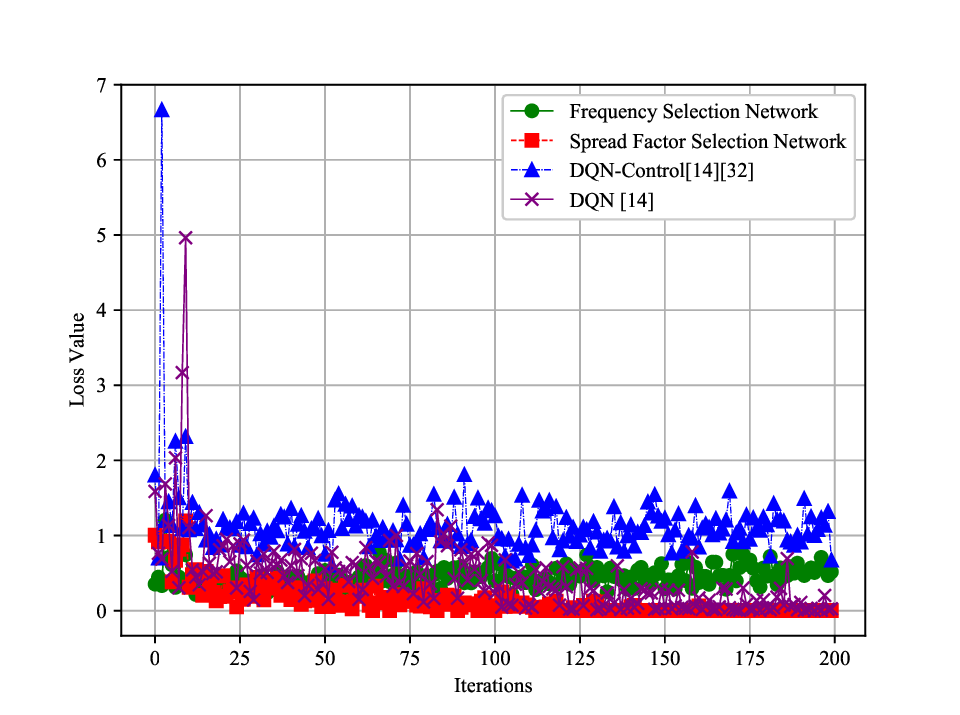}
  \caption{The loss value of different algorithms.}\label{Fig:Loss}
\end{center}
\end{figure}
In this subsection, the following comparisons were primarily conducted: First, a performance comparison between three algorithms and the proposed algorithm in the aforementioned scenario. Second, a performance comparison under different learning rates. Third, a performance comparison of the algorithm under different action ranges. Fourth, a performance comparison of the algorithm under different channel models.
In Fig.~\ref{Fig:Result}, four algorithms are compared. The first is our algorithm with a $\varepsilon$-greedy mechanism to validate the effectiveness of our design after removing the $\varepsilon$-greedy mechanism. The second approach employs the DQN algorithm for frequency selection~\cite{8314744}, while the control of spreading factors is guided by the principles outlined in~\cite{10120634} to validate the effectiveness of the interconnected reward function design for enhanced coordination of selections. Specifically, the adjustment of spreading factors is based on the magnitude of the feedback SJNR. If the SJNR significantly exceeds the required level, the spreading factor is reduced to minimize spectrum occupancy. Conversely, if the SJNR falls below a predetermined threshold, the spreading factor is increased to enhance anti-jamming performance and thereby achieve better communication robustness. The third algorithm is a DQN-based control method that manages both frequency and spreading factor~\cite{8314744} to validate the effectiveness of the parallel network architecture.
\begin{figure}[h]
\begin{center}
  \vspace{-1mm}
  \includegraphics[width=3.5in]{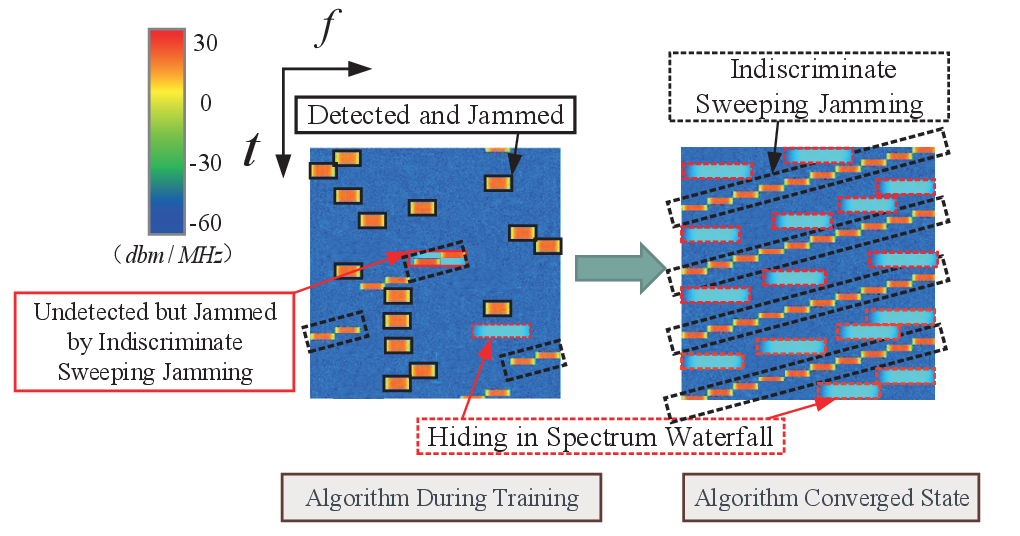}
  \caption{The Comparison of spectrum waterfalls during algorithm training and converged state.}\label{Fig:Illustration}
\end{center}
\end{figure}

\begin{figure}[h]
\begin{center}
  \vspace{-1mm}
  \includegraphics[width=3.5in]{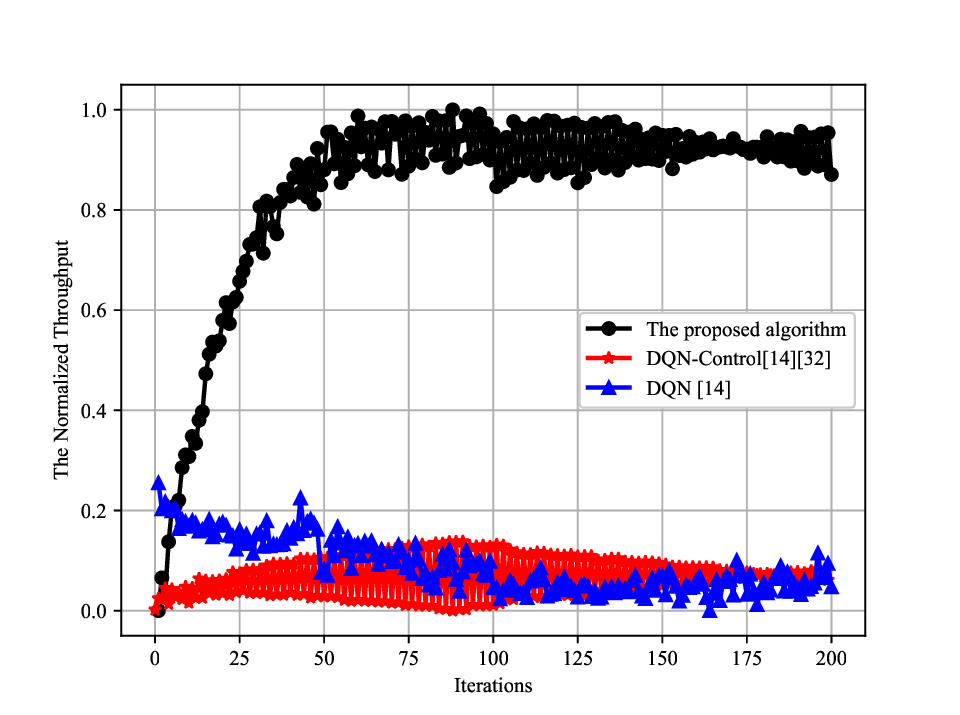}
  \caption{The performance under larger spread factor set.}\label{Fig:SpreadFactor}
\end{center}
\end{figure}
\begin{figure}[h]
\begin{center}
  \vspace{-1mm}
  \includegraphics[width=3.5in]{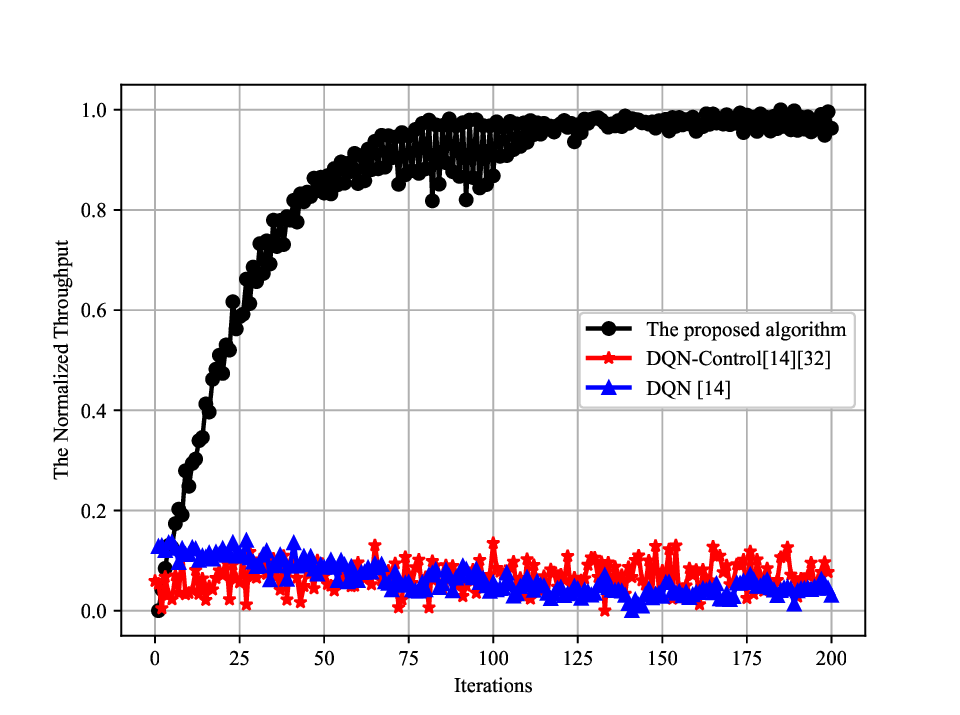}
  \caption{The performance under shadow fading channel.}\label{Fig:shadow}
\end{center}
\end{figure}

In Fig.~\ref{Fig:Result}, the horizontal axis represents the algorithm iterations, while the vertical axis denotes the normalized throughput. \textcolor[rgb]{0000,0.00,0000}{ Normalized throughput is defined as the ratio of the actual amount of successfully transmitted data to the maximum possible data that could be transmitted under that jamming condition which it can be expressed as:
\begin{equation}\label{eq:Thought}
{C_{Normal}} = \frac{{{C_{Actual}}}}{{{C_{Max}}}}
\end{equation}
where $C_{Actual}$ represents the throughput achieved with the parameters selected by the algorithm at a given time, $C_{Max}$ represents the maximum throughput that could be achieved under the same jamming condition, determined through exhaustive search or optimal parameter selection.
}
Evidently, both the DQN algorithm~\cite{8314744} and the DQN-based control algorithm~\cite{10120634} exhibit poor performance. The underlying issue with the conventional DQN approach lies in the curse of dimensionality, which leads to a vast action space, hindering the algorithm's convergence~\cite{henderson2018deep}. Meanwhile, the DQN-based control algorithm struggles to effectively integrate frequency selection with spreading factor control. However, in the proposed algorithm, the dimensionality problem is mitigated through action decomposition by two networks. Compared with our designed algorithm that eliminates the $\varepsilon$-greedy mechanism, this not only accelerates the convergence of our algorithm but also obviates the need to tune the $\varepsilon$-greedy parameter. To address the challenge of joint decision-making, our design incorporates a correlated reward mechanism in the loss functions of two interconnected networks. This sophisticated arrangement enables our algorithm to excel at coordinated decision-making processes, thereby overcoming the limitations observed in the compared methodologies. To keep the simulation images concise, we validate the performance of our proposed algorithm and two other algorithms in other scenarios.

In Fig.~\ref{Fig:Loss}, the horizontal axis represents the iterations of the algorithm, while the vertical axis denotes the loss values. Evidently, both the frequency selection network and the spread spectrum network exhibit low and stable loss values, with the frequency selection network showing a marginally higher loss compared to the spread spectrum factor selection network. Conversely, the DQN, particularly within the context of the aforementioned spread spectrum factor control algorithms, yields higher loss values accompanied by significant fluctuations, indicative of an algorithm that has not yet reached convergence.

In Fig.~\ref{Fig:Illustration}, it can be seen that when the algorithm is in training, an incorrect selection of spreading factor can easily be detected by the jammer and jammed. Even if the correct spreading factor is chosen, selecting an inappropriate frequency point can also lead to jam from indiscriminate jamming signals. After the algorithm converges, the joint selection of frequency and spreading factor can be observed from the spectrum waterfall. The proposed algorithm can choose the appropriate spreading factor to avoid detection by the jammer while selecting suitable frequencies to avoid being swept by jamming signals, thereby increasing overall throughput.
\begin{figure*}[h]
\begin{center}
  \vspace{-1mm}
  \includegraphics[width=7in]{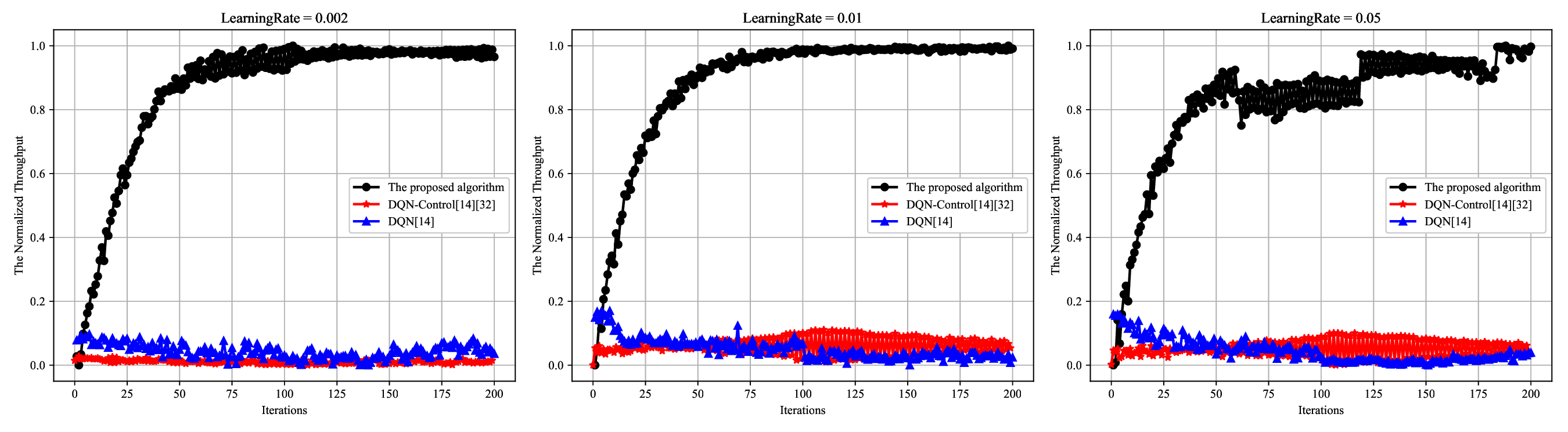}
  \caption{The performance comparison of different learning rates.}\label{Fig:learningRate}
\end{center}
\end{figure*}
For direct sequence spread spectrum (DSSS) systems~\cite{10561514}, spreading factors often exceed 128. To accommodate potential future hardware advancements, we tested higher spreading factors with frequency hopping capabilities, thus setting the spreading factors from \{1,2,4,8,16,32\} to \{1,2,4,8,16,32,64,128,256\}. As Fig.~\ref{Fig:SpreadFactor} illustrates, as the number of spreading factors increases, there are perturbations in the algorithm's convergence. However, the proposed algorithm still outperforms the comparison algorithms in terms of performance.

In Fig.~\ref{Fig:shadow}, the channel model is changed to consider shadow fading~\cite{995511}. With shadow fading, the proposed algorithm has fluctuations, but it can still maintain much better performance than other algorithms. Comparing Fig.~\ref{Fig:SpreadFactor} with Fig.~\ref{Fig:shadow}, a larger action space has a greater impact on the algorithm's convergence.

\textcolor[rgb]{0000,0.00,0000}{In Fig.~\ref{Fig:learningRate}, the performance is compared against learning rates of 0.002, 0.01, and 0.05. }It is evident that, under various hyperparameter settings, the algorithm maintains commendable performance, with the exception of when the learning rate is set to 0.05, where performance begins to fluctuate. According to the findings from~\cite{henderson2018deep}, an excessively high learning rate indeed impacts the convergence properties of the algorithm. Therefore, when applying this algorithm in practice, it is recommended that the learning rate be kept below 0.05 to ensure stable and effective performance.

\begin{figure}[h]
\begin{center}
  \vspace{-1mm}
  \includegraphics[width=3.5in]{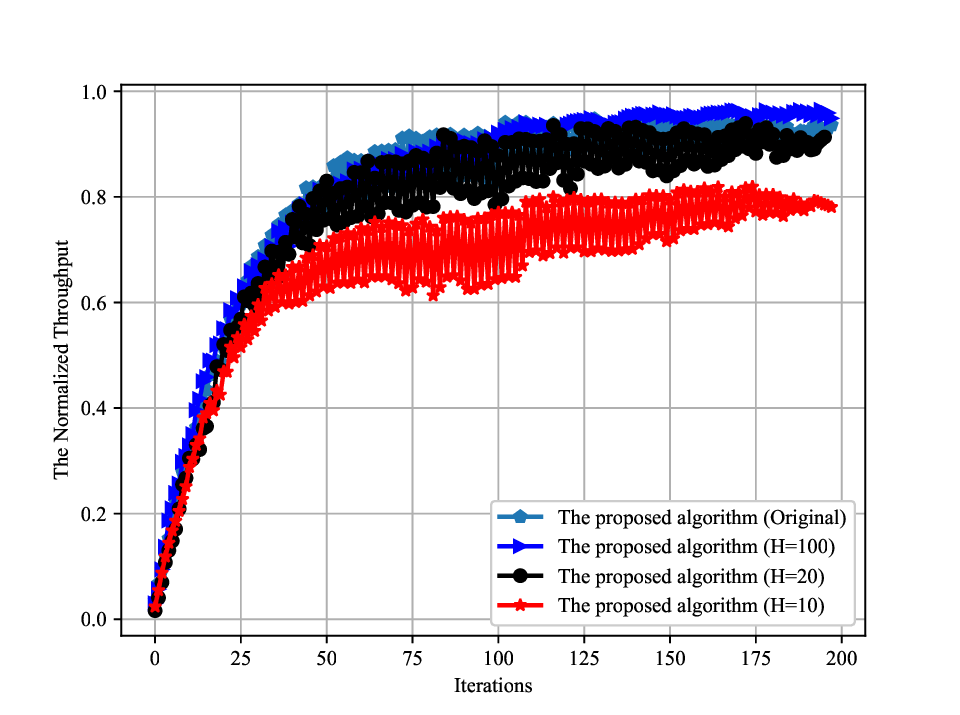}
  \caption{The performance in constant jamming.}\label{Fig:DifferentH}
\end{center}
\end{figure}
\textcolor[rgb]{0000,0.00,0000}{
In proof, the retrospective length $H$ is important in MDP. The retrospective length
$H$ can be regarded as the length of the spectrum waterfall. As illustrated in Fig.~\ref{Fig:DifferentH}, a smaller $H$ value leads to a more pronounced decline in performance. This decline occurs because, with an excessively small $H$, the proposed algorithm struggles to effectively account for the impact of past actions on the current state. Specifically, it becomes difficult for the algorithm to recognize the frequency sweeping patterns after selecting an appropriate spreading factor, which in turn makes it challenging to evade such interference and results in performance degradation. However, the algorithm can still adapt based on its previous action selections, enabling it to choose suitable spreading factors and frequencies that offer better evasion opportunities. Consequently, there remains a probability of successfully evading frequency sweeping jamming. Simulation results reveal that at $H$ = 20, the algorithm can already largely identify the frequency sweeping patterns, ensuring that performance remains relatively unaffected at this retrospective length.}

\begin{figure}[h]
\begin{center}
  \vspace{-1mm}
  \includegraphics[width=3.5in]{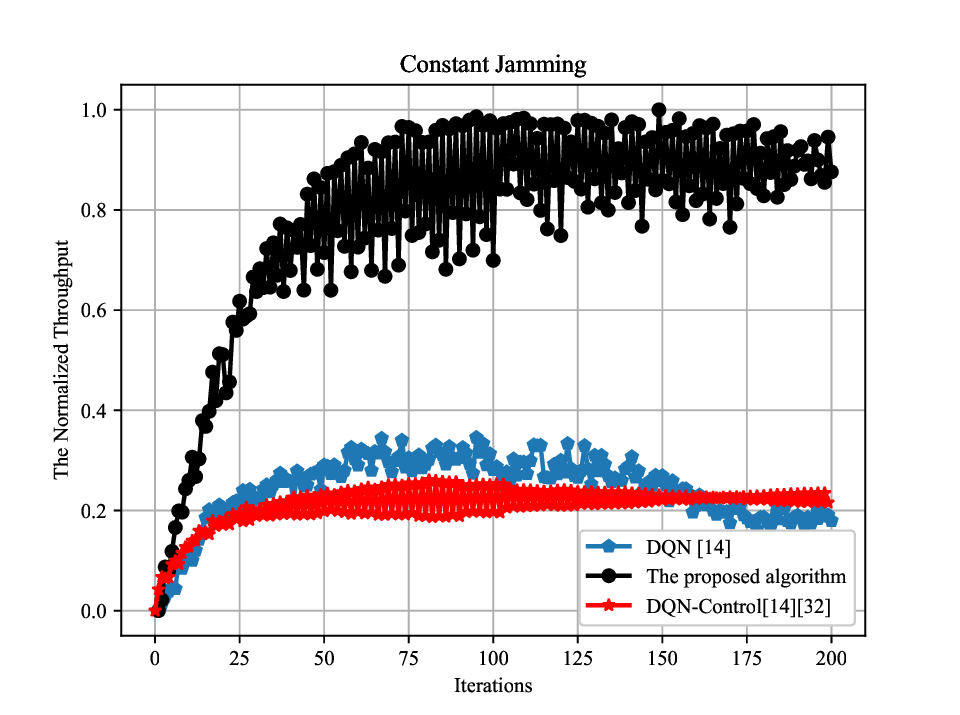}
  \caption{The performance in constant jamming}\label{Fig:con}
\end{center}
\end{figure}

\begin{figure}[h]
\begin{center}
  \vspace{-1mm}
  \includegraphics[width=3.5in]{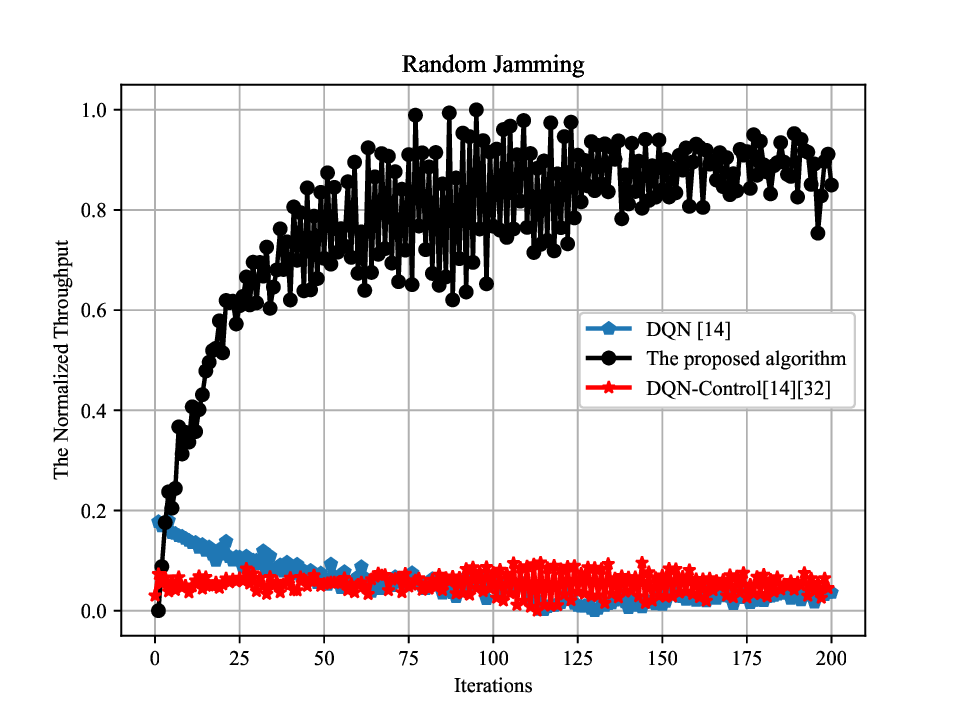}
  \caption{The performance in random jamming.}\label{Fig:Random}
\end{center}
\end{figure}

\textcolor[rgb]{0000,0.00,0000}{As illustrated in Fig.~\ref{Fig:con} and~Fig.~\ref{Fig:Random}, two jamming scenarios to test performace of our algorithms: constant jamming~\cite{9733393}, characterized by emissions over a 10 MHz bandwidth without prior signal detection, and random jamming~\cite{9733393}, which is limited to a 2 MHz bandwidth. The simulation results demonstrate that the proposed algorithm retains its effectiveness across both jamming conditions. Notably, under constant jamming, where the emission pattern is more predictable, an improvement in the performance of the comparative algorithms is observed.}

\subsection{Time Complexity}

Based on the parameters in Table~\ref{tab:Parameters}, the computational requirements of the deep networks in the simulation can be roughly estimated.
The FLOPs (Floating Point Operations Per Second) of the CNN for one time step can be roughly estimated as~\cite{He_2015_CVPR}:
\begin{equation}\label{eq:cnn}
\begin{split}
{FLOPs}_{{CNN}} = &2 \times {kernelSize}^2 \times {inputChannels} \\
&\times {outputWidth} \times {outputHeight} \\
&\times {outputChannels},
\end{split}
\end{equation}
where input channels and output channels refer to the distinct components of the initial data, like the red, green, and blue channels in an RGB image. \textcolor[rgb]{0000,0.00,0000}{Due to the challenge of high computational demands when processing RGB images, we convert them into grayscale images~\cite{5445596}. For grayscale image, the input channel and output channel are both 1.}
The FLOPs of fully connected layers for a single time step is given by~\cite{goodfellow2016deep}:
\begin{equation}\label{eq:fc}
FLOPs_{fc} = inputNeurons \times outputNeurons \times 2,
\end{equation}
where $inputNeurons$ and $outputNeurons$ refer to the sizes of both the input and output layers as depicted in Table~\ref{tab:Parameters}.

In summary, based on \textcolor[rgb]{0000,0.00,0000}{Eq.~\ref{eq:cnn} and Eq.~\ref{eq:fc}} and the data provided in Tab.~\ref{tab:Parameters}, the computational cost of the proposed algorithm can be calculated to be around $3.3 \times 10^9$. According to~\cite{10528752}, the FLOPs of a CPU are around $10^{12}$, indicating that the proposed algorithm is compatible with many hardware configurations, such as Jetson TX2 and Jetson AGX~\cite{9152915}.

\begin{table}[htbp]
\centering

\caption{Version of packages}
\label{tab:packages}
\begin{tabular}{ll}
\hline
\textbf{Library Name} & \textbf{Version} \\
PyTorch               & 2.1.0            \\
TorchVision           & 0.16.0           \\
NumPy                 & 1.26.0           \\
SciPy                 & 1.11.3           \\
Pandas                & 2.1.1            \\
Matplotlib            & 3.8.0            \\
OpenCV                & 4.2.0            \\ \hline
\end{tabular}
\end{table}

\begin{table}[htp]
\centering

\caption{Network Parameters for Compared Algorithms}
\label{tab:networkC}
\begin{tabular}{|c|c|c|c|c|}
\hline
\textbf{Algorithm} &
  \textbf{Layer} &
  \textbf{Setting} &
  \textbf{Input} &
  \textbf{Output} \\ \hline
\multirow{5}{*}{\begin{tabular}[c]{@{}c@{}}DQN-\\ Control\end{tabular}} &
  \begin{tabular}[c]{@{}c@{}}Convolutional\\ layer\end{tabular} &
  \begin{tabular}[c]{@{}c@{}}kernel size=4, \\ stride=2\end{tabular} &
  200*200 &
  99*99 \\ \cline{2-5}
 &
  \begin{tabular}[c]{@{}c@{}}Convolutional\\ layer\end{tabular} &
  \begin{tabular}[c]{@{}c@{}}kernel size=4, \\ stride=2\end{tabular} &
  99*99 &
  48*48 \\ \cline{2-5}
 &
  \begin{tabular}[c]{@{}c@{}}Fully connect\\ layer\end{tabular} &
  \begin{tabular}[c]{@{}c@{}}Activation: \\ ReLU\end{tabular} &
  48*48 &
  512 \\ \cline{2-5}
 &
  \begin{tabular}[c]{@{}c@{}}Fully connect\\ layer\end{tabular} &
  \begin{tabular}[c]{@{}c@{}}Activation: \\ ReLU\end{tabular} &
  512 &
  256 \\ \cline{2-5}
 &
  \begin{tabular}[c]{@{}c@{}}Fully connect\\ layer\end{tabular} &
  \begin{tabular}[c]{@{}c@{}}Activation: \\ ReLU\end{tabular} &
  256 &
  10 \\ \hline
\multirow{5}{*}{DQN} &
  \begin{tabular}[c]{@{}c@{}}Convolutional\\ layer\end{tabular} &
  \begin{tabular}[c]{@{}c@{}}kernel size=4, \\ stride=2\end{tabular} &
  200*200 &
  99*99 \\ \cline{2-5}
 &
  \begin{tabular}[c]{@{}c@{}}Convolutional\\ layer\end{tabular} &
  \begin{tabular}[c]{@{}c@{}}kernel size=4, \\ stride=2\end{tabular} &
  99*99 &
  48*48 \\ \cline{2-5}
 &
  \begin{tabular}[c]{@{}c@{}}Fully connect\\ layer\end{tabular} &
  \begin{tabular}[c]{@{}c@{}}Activation: \\ ReLU\end{tabular} &
  48*48 &
  512 \\ \cline{2-5}
 &
  \begin{tabular}[c]{@{}c@{}}Fully connect\\ layer\end{tabular} &
  \begin{tabular}[c]{@{}c@{}}Activation: \\ ReLU\end{tabular} &
  512 &
  256 \\ \cline{2-5}
 &
  \begin{tabular}[c]{@{}c@{}}Fully connect\\ layer\end{tabular} &
  \begin{tabular}[c]{@{}c@{}}Activation: \\ ReLU\end{tabular} &
  256 &
  10*6 \\ \hline
\end{tabular}
\end{table}

\begin{table}[htbp]
\centering

\caption{Runtime of algoritms}
\label{tab:time}
\begin{tabular}{ll}
\hline
\textbf{Algorithm} & \textbf{Run time for one iteration (second)} \\ \hline
DQN                & 0.00026815                                   \\
DQN-Control        & 0.00026763                                    \\
Ours               & 0.00027991                                    \\ \hline
\end{tabular}
\end{table}
\textcolor[rgb]{0000,0.00,0000}{In addition, we evaluated the runtime of these algorithms. The testing platform specifications include an operating system of Ubuntu 22.04, a CPU of i9-13900K, and a GPU of RTX 4080 Ti. Refer to Table~\ref{tab:packages} for the versions of other packages. The network parameter of compared algorithm is shown in Table~\ref{tab:networkC}. The runtimes of each algorithm are summarized in Table~\ref{tab:time}, where the reported times are averages over multiple simulations. In a single iteration, the proposed algorithm adds approximately 0.0001 seconds to the runtime compared to the others. As noted in~\cite{He_2015_CVPR}, the combination of CNN and FCNN networks primarily concentrate computational resources on convolution operations. In contrast, the proposed parallel network architecture utilizes fully connected neural networks, which explains the minimal increase in runtime.
}

\section{Conclusion}

In conclusion, this paper has explored the critical issue of anti-jamming under the threat of moving reactive jammers. Through the innovative design of a parallel DRL algorithm, we addressed the dimensionality curse associated with multi-decision anti-jamming strategies in the presence of \textcolor[rgb]{0000,0.00,0000}{moving reactive jammers}. By leveraging the parallel network structure for action space decomposition, the algorithm significantly reduces the complexity of decision-making, thereby overcoming the challenges imposed by high-dimensional action spaces. Furthermore, we have harnessed the inherent randomness within the DRL framework to eliminate the conventional $\varepsilon$-greedy mechanism, thus refining the balance between exploration and exploitation. This modification has led to a notable acceleration in the convergence of our proposed algorithm, demonstrating its efficiency and adaptability in dynamic and adversarial environments. Simulations have substantiated the efficacy of our approach. Notably, in comparisons measuring normalized throughput, the performance enhancements achieved by our algorithm exceeded 90\%, highlighting its superiority over baseline methods.

\bibliographystyle{IEEEtran}
\bibliography{reference}
\end{document}